\documentclass[12pt]{article}

\usepackage{arxiv}
 
\usepackage{setspace}

\usepackage[utf8]{inputenc} % allow utf-8 input
\usepackage[T1]{fontenc}    % use 8-bit T1 fonts
\usepackage{hyperref}       % hyperlinks
\usepackage{url}            % simple URL typesetting
\usepackage{booktabs}       % professional-quality tables
\usepackage{amsfonts}       % blackboard math symbols
\usepackage{nicefrac}       % compact symbols for 1/2, etc.
\usepackage{microtype}      % microtypography
\usepackage{lipsum}

\usepackage{caption}
\usepackage{graphicx}
\usepackage{lscape}
\usepackage{hyperref}
\usepackage{amsmath}
\usepackage{longtable}%
\usepackage{colortbl}%

\title{Does Crowdfunding Really Foster Innovation?\\Evidence from the Board Game Industry}

\author{
  Johannes Wachs\\
  Vienna University of Economics and Business\\
  Complexity Science Hub Vienna\\
  \texttt{johannes.wachs@wu.ac.at} \\
   \And
 Bal\'azs Vedres \\
  University of Oxford\\
  Central European University\\
    \texttt{balazs.vedres@oii.ox.ac.uk} \\
}

\begin{document}
\maketitle

\begin{abstract}
Crowdfunding offers inventors and entrepreneurs alternative access to resources with which they can develop and realize their ideas. Besides helping to secure capital, crowdfunding also connects creators with engaged early supporters who provide public feedback. But does this process foster truly innovative outcomes? Does the proliferation of crowdfunding in an industry make it more innovative overall? Prior studies investigating the link between crowdfunding and innovation do not compare traditional and crowdfunded products and so while claims that crowdfunding supports innovation are theoretically sound, they lack empirical backing. We address this gap using a unique dataset of board games, an industry with significant crowdfunding activity in recent years. Each game is described by how it combines fundamental mechanisms such as dice-rolling, negotiation, and resource-management, from which we develop quantitative measures of innovation in game design. Using these measures to compare games, we find that crowdfunded games tend to be more distinctive from previous games than their traditionally published counterparts. They are also significantly more likely to implement novel combinations of mechanisms. Crowdfunded games are not just transient experiments: subsequent games imitate their novel ideas. These results hold in regression models controlling for game and designer-level confounders. Our findings demonstrate that the innovative potential of crowdfunding goes beyond individual products to entire industries, as new ideas spill over to traditionally funded products.

\keywords{Crowdfunding, innovation, novelty, board games, entrepreneurship}
\end{abstract}

\section*{Introduction}
Entrepreneurs are increasingly turning to crowdfunding to raise capital for their ideas, and to involve an interested public in the process of creation. Crowdfunding platforms offer a large scale and decentralized alternative to traditional sources of capital. One distinguished example of a crowdfunded product is the Oculus Rift virtual reality headset, which was eventually acquired by Facebook for over \$$1$ billion. While previous research documents key differences in how decentralized crowds and experts decide who or what to fund \cite{stanko2017toward,kuppuswamy2017does,petruzzelli2019understanding}, less is known about how products created through crowdfunding efforts are distinct from their traditionally funded counterparts.

Crowdfunding is often framed as a potential antidote to specific problems in traditional industry. Gatekeeping is one such issue thought to harm innovation in creative industries. The problem of gatekeeping refers to the notion that, historically, decisions about the allocation of capital have been up to a small, homogeneous slice of society \cite{younkin2016problems}. Members of this decision-making group may naturally be drawn to conventional alternatives by strong social forces such as homophily \cite{greenberg2015leaning}, familiarity bias, or risk aversion. Such preferences can manifest in the exclusion of newcomers or even established entrepreneuers with innovative ideas \cite{mollick2016democratizing}. Even for capital intensive ideas, a small-scale prototype created with funds from the crowd can open doors \cite{kaminski2019new}. From this perspective online crowdfunding, as a technology, represents a shift from hierarchical to decentralized  modes of capital allocation \cite{malone1987electronic}. 

Previous work also highlights other ways in which crowdfunding supports ideas that may otherwise languish \cite{younkin2016problems}, for instance by facilitating coordination with end-users and communication with supporters \cite{herve2018crowdfunding}. Such interactions are known to be an important source of ideas and support in the creative process of bringing a product to market \cite{stanko2017toward,clauss2018directing}. Another benefit of a supportive crowd is that they double as emotionally-invested early adopters \cite{agrawal2014some}, seeding potential network effects in product adoption down the line and lending valuable credibility to nascent projects \cite{kaminski2019new,alt2019electronic}.  

So while we do know from prior work that crowdfunding offers a real alternative to traditional funding, and that it both expands and shapes the field of products that are launched, we do not know if the end products of crowdfunding efforts are really different from traditionally funded ones. It is conceivable that the process of crowdfunding tends towards stereo-typical solutions or chases fads, binding the entrepreneur to the lowest common denominator of tastes in a group. The crowd may steer an entrepreneur towards increasingly niche ideas \cite{stevenson2019out} or exhibit strong biases against entrepreneurs that are members of racial minorities \cite{younkin2018colorblind}. With these competing factors in the background, evaluating the innovation outcomes of crowdfunded products is important because public support for and general interest in crowdfunding generally assumes that it increases innovation \cite{valanciene2013valuation}. 

In this paper we address this question using a novel dataset of nearly all board games published in recent decades sourced from BoardGameGeek \footnote{See: \url{https://boardgamegeek.com/}} (BGG), an online portal curated by thousands of hobby gamers. Unlike previous empirical studies, which tend to compare crowdfunded products amongst themselves, our dataset contains many examples of both crowdfunded and traditionally published games. The proportion of crowdfunded projects in our database increased from practically none in 2006 to more than 30 percent in 2017. This case offers a unique opportunity to analyze the impact of crowdfunding on the nature of innovation for an entire industry. 

To compare different games, we characterize them by the combination of mechanisms they incorporate. The BGG community maintains a list of 51 mechanisms which games can have, for example dice rolling, pattern recognition, negotiation, and player elimination. Some games are simple: the famous children's game \textit{Candyland} has only the ``roll/spin and move'' mechanism. Others are much more complex, sometimes combining over ten mechanisms. These features place games in a 51 dimensional feature space. We emphasize that it is the combination of these fundamental building blocks that describe the essence of games and differences between them \cite{elverdam2007game}. Recent studies of creativity and innovation in diverse cultural products including music \cite{askin2017makes}, video games \cite{de2015game}, and political speech \cite{barron2018individuals} use a similar multi-dimensional perspective. Distances such spaces encode distinctiveness and novelty of products relative to their peers.

Using this embedding of games into the space of their mechanisms, we observe several differences between crowdfunded and traditional games using a regression modeling framework. Compared to previously published games, a crowdfunded game tends to be more \textit{distinct} than similarly aged traditional games, measured by their distance in the mechanism space. Crowdfunded games are also more \textit{novel}, in that they are more likely to combine pairs of mechanisms that have not been combined before. This suggests that crowdfunded innovation is often substantive rather than marginal. Lastly, we observe that future games are more similar to crowdfunded games than traditional games, suggesting that crowdfunded innovation is \textit{resonant}, setting trends that others follow and shifting the direction of the entire industry.

The rest the paper is structured as follows. First we review theoretical arguments for the potential of crowdfunding as an accelerant for innovation. Then we introduce the board game industry and our dataset from which we derive measures of innovation and other features. To help conceptualize our core message, we visualize the landscape of board game types at various points in time as a network, highlighting hotspots of crowdfunding. We then present our models and interpret the results. We conclude with a discussion of our work, its limitations, and potential extensions and future work.

%In many industries crowdfunding offers creators and inventors of new products an alternative market for capital. Firms and financial institutions that provide capital to entrepreneurs represent hierarchical modes of organization which are better capable of complex coordination [cite Oliver Williamson paradigm] needed to support new products. As these costs of coordination are falling with advances in ICT, crowdfunding has become an increasingly attractive alternative. The resulting proliferation of crowdfunding can be viewed as a shift from relatively hierarchical to a more market oriented mode of capital allocation \cite{malone1987electronic}.

\section*{The Innovative Potential of Crowdfunding}

Crowdfunding has emerged as a significant institution for raising capital for new ventures. It has been involved in a number of highly visible success stories, such as the Oculus Rift VR headset, the Pebble smartwatch, a variety of 3D printers, and countless other projects from games to works of art. Crowdfunding, we argue, is not only an avenue to access funding, but it is also an organizational innovation that fosters creativity \cite{testa2019role}. Crowdfunding combines aspects of traditional financial intermediation with an expert public. These two forces mutually transform each other: financing becomes a public testament of quality, by virtue of the expert public watching money pledged, and opinions voiced are made credible by the funding committed, serving as an entry ticket to the expert public. Funders offer both a commitment of money and usually a public form of support. Successful campaigns win legitimacy \cite{soubliere2020legitimacy}. Serial entreprenuers build loyal networks brimming with social capital \cite{skirnevskiy2017influence} and embedded trust \cite{horvat2015network}. 

Though there are historical examples of crowdfunding, for instance by the early social scientist and philosopher Auguste Comte\cite{galuszka2017crowdfunding}, who solicited contributions from the public to support his work, as an institution crowdfunding is inextricably tied to the web. It is only as an electronic market that crowdfunding could realize its potential as a tool for both outreach and coordination. The intensely public nature of digital crowdfunding also creates important externalities: competitors surely observe and learn from the ideas and experiences of the crowdfunding entrepreneur \cite{agrawal2014some}.

Crowdfunding is an institution that diversifies the base of potential entrepreneurs. Through engagement with the crowd and the pressures of public visibility, it exerts substantial force on the creation of new products. It is a fundamentally digital institution because it is only through the scale and public nature of the web that it can serve as an effective and reliable market. Crowdfunding decreases an entrepreneur's uncertainty about her product through interactions with potential customers and observations of competitors. It also decreases uncertainty for the people that deal with the entrepreneur: potential investors, partners, collaborators can observe how she works and how others receive her work \cite{podolny2001networks}.

We claim that these institutional features of crowdfunding facilitate innovation. It opens a path for outsiders to raise funds, enables creators to iterate on high risk ideas with tight feedback loops, and diffuses the risk of supporting novel ideas among the crowd. We therefore hypothesize that, keeping features of creators and products constant, \textbf{crowdfunded products will be more distinctive than traditionally published products}. Beyond distinctiveness or atypicality, we also argue that \textbf{crowdfunded products will more often manifest some entirely novel idea or component}. This is an important distinction - crowdfunded products may seem relatively distinct if they are quickly copying the most innovative products created by traditional means. We will argue and demonstrate that this is not the case: crowdfunding is not simply a tool that facilitates imitation and iteration on the latest inventions. Our last hypothesis about crowdfunding is that \textbf{crowdfunded products are influential, inspiring imitation by subsequently published games.} Crowdfunding happens in the open and competitors can observe and learn from crowdfunded projects. In this way the iteration that happens between entrepreneur and crowd not only creates more appealing projects and quality innovation, but can facilitate imitation by other designers \cite{hui2014understanding}. We will test these hypotheses using data from the board game industry, which we now describe.

\subsection*{The Board Game Industry}
Board games have a rich social and anthropological history \cite{woods2012eurogames}. Games such as Monopoly and Scrabble are cultural icons of post-World War 2 consumer society. A market for more sophisticated games emerged in the mid-1990s, when a nascent style of so-called ``eurogames'' flourished. These complex games eschewed the simplistic design of games such as Monopoly. A primary example of this style is Klaus Teuber's \textit{Settler's of Catan} a game published in 1995 which has sold over 25 million copies worldwide. In recent years, the board games industry has undergone a boom\footnote{For an overview from the popular press, see: \url{https://www.nytimes.com/2019/09/02/business/crowdfunding-board-game-inventors.html}}, with Statista estimating global revenues in 2018 at around \$12 billion. Several of the most popular games of the last decade have been crowdfunded, including \textit{Cards Against Humanity} and \textit{Gloomhaven}, the top ranked game on BoardGameGeek. 

Board games may be especially amenable to crowdfunding \cite{tyni2020double} and the case of the board game industry an especially suitable natural laboratory to understand how crowdfunding can contribute to the innovativeness of an entire field. Games have relatively low capital requirements and can be conceptualized and designed by individuals. The coordination overhead to develop and distribute a board game is substantially less complex than that required, for example, to develop a video game or medical devices. Board game development can also take make good use of feedback from the crowd: designers often emphasize the importance of user-testing in the development process. For instance Matt Leacock, designer of the best-selling \textit{Pandemic} series of games, credits early play-testers with the idea to add a new mechanism to the first game of the series \cite{leacockVideo}. Leacock's background as a user experience designer suggests that he was aware of the value of gathering feedback from potential users and iterating on their inputs. The crowdfunding model connects entrepreneurs directly with an interested audience, facilitating this important kind of feedback \cite{von2006democratizing}.

Specific examples from the early history of Kickstarter highlight other ways the crowdfunding process can help bring different kinds of products to market. One of the first crowdfunded hits, \textit{Alien Frontiers}, the fourth ever board game project on Kickstarter raised nearly \$15,000 in 2010 \cite{morganHow}. Tory Niemann, the game's designer, had unsuccessfully pitched previous ideas to mainstream publishers. As an outsider, he turned to crowdfunding to get around traditional gatekeepers. A few months later, another game called \textit{Eminent Domain} raised nearly \$50,000. The team behind this title had previous experience published games with traditional publishers, but was interested in the potential interactions with end-users that they had seen the designer of \textit{Alien Frontiers} benefit from \cite{jaffeeInterview}. That the team behind \textit{Eminent Domain} was aware of the process of crowdfunding \textit{Alien Frontiers} went through highlights the potential for knowledge spillovers in crowdfunding. These two games signaled the start of a significant acceleration in crowdfunding in the board game industry, powered by platforms including Kickstarter and Indiegogo.

%The costs of creating a prototype board game are relatively low, and most games are designed by individuals or small groups. As an example, in Figure~\ref{fig:pandemic_evolution} we show the best-seller \textit{Pandemic} by designer Matt Leacock in three stages of development, from concept diagram with notes on paper, to first prototype created in a graphic design suite, to final product created in collaboration with a publishing house and released in 2008 \cite{leacockVideo}. Pandemic is a cooperative game in which players work together to contain the spread of infectious diseases on a map of the world. According to the Board Game Geek typology, \textit{Pandemic} has seven mechanisms: ``Action Point Allowance System'', ``Co-operative Play'', ``Hand Management'', ``Point to Point Movement'', ``Set Collection'', ``Trading'', ``Variable Player Powers''. This example illustrates that board games are highly conceptual products that can be developed by small groups with relatively little capital investment, making this industry perhaps an ideal one for crowdfunding. Leacock also notes that while many of the mechanisms in his game were defined in his earliest prototype, user-testing at board game expos and private parties led to important tweaks and improvements, including for instance the ``Variable Player Powers'' mechanism \cite{leacockVideo}. This highlights the potential value of the crowd as a source of creative input.

% mechanism list. on landscape. how innovation is measured
% text for pandemic, AF/ED, CAH

\section*{Data and Methods}
We collected data from BoardGameGeek (BGG), an online database containing tens of thousands of board games. Created in 2000, BoardGameGeek has emerged as the online hub of board game enthusiasts. Besides storing information about board games, the site has a lively web forum and organizes real life conferences that have become an important part of the board game industry. It boasts over a million registered users and an Alexa ranking in the range of 2000-3000. Its role in the industry can be compared with the role that IMDB or Rotten Tomatoes plays for the film industry. Game creators have significant incentive to add their game to BGG.

We collected data on all games on the website by early December 2017. From each game's page we extracted a variety of features that describe the game. These include traditional descriptive attributes of the games, such as their genre, recommended minimum age to play, and characteristic playing time, and feedback from users about the game, including their perceived quality (a rating on a 1-10 scale) and ``weight'' or complexity (a rating on a 1-5 scale). Games that were crowdfunded are labeled. 

BoardGameGeek maintains a curated list of 51 mechanisms that describe the core functional elements present games. Examples include \textit{dice-rolling}, \textit{hand management}, and \textit{player elimination}. We list the full set of mechanisms in the Appendix (see Table \ref{Appendix:mechanism_table}). Each game's page lists its mechanisms. Using them as a description of the game, we abstract games as 51-dimensional binary vectors. Though mechanisms themselves do not capture all aspects of gameplay~\cite{kritz2017building}, much of what makes games unique is how their mechanisms interact. We use this description of games to compare the relative position of crowdfunded and traditional games in the 51-dimensional mechanism space, and to study how these positions evolve over time.

We filter the data to make reasonable comparisons between crowdfunded and traditionally funded games. We consider games which list a designer, and have at least two mechanisms. To reduce the likelihood that we analyze games with missing or inaccurate metadata, we consider only those games rated by at least 10 BGG users. Finally, we filter out trivial expansions: those expansions of previously made games which do not change their mechanisms. For instance, we do not want to overestimate the novelty of crowdfunded games by comparing them to the hundreds of rebranded/reissued versions of Monopoly (i.e. Monopoly London), which are more common among traditional games. After these steps, we are left with 9,170 games published since 2006.

\begin{figure}[!ht]
\centering
\includegraphics[width=.8\textwidth]{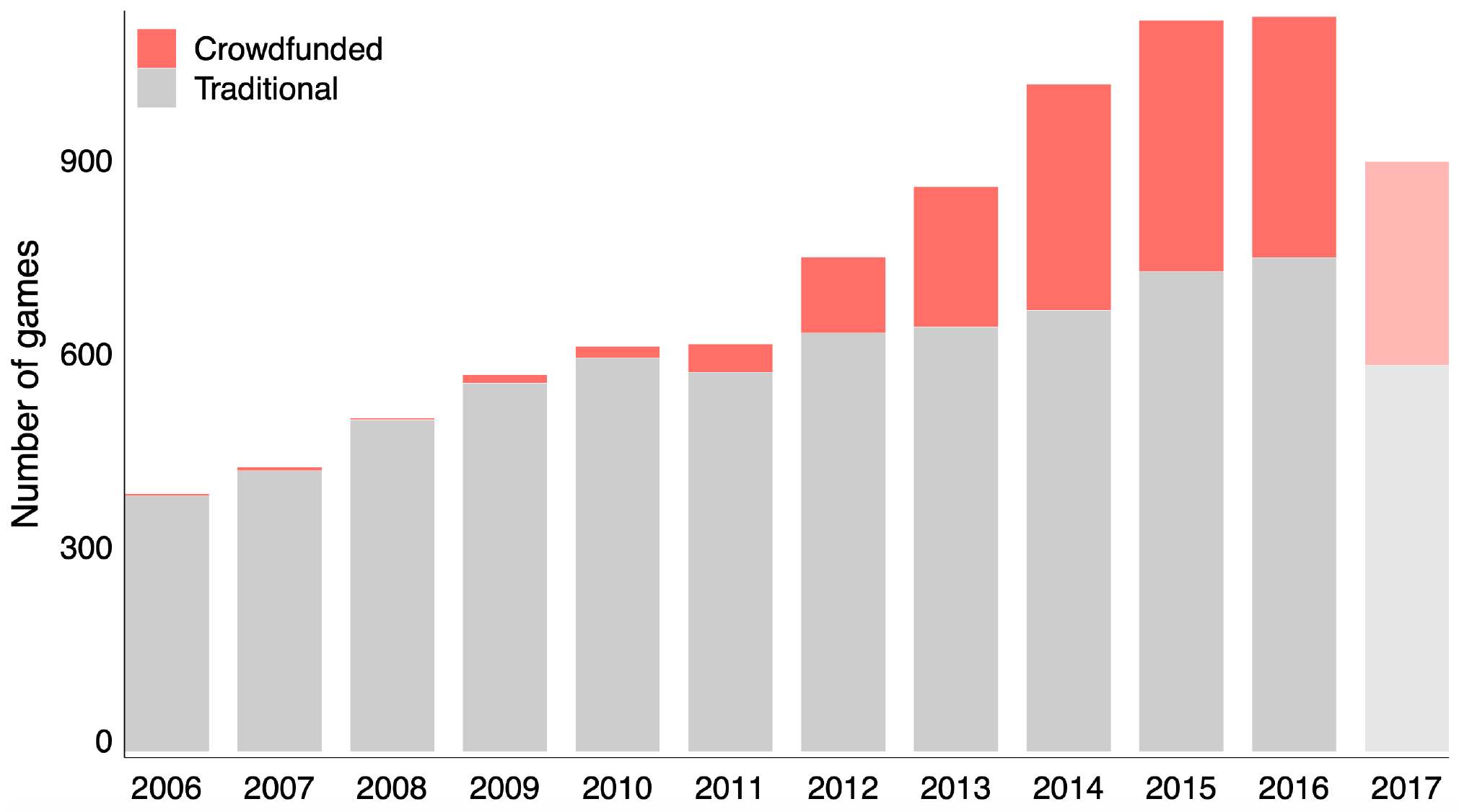}
\caption{Count of traditional and crowdfunded games by year. Data was collected during 2017, and does not include all games published in that year. The share of crowdfunded games in our data rises from below 1\% to around 30\% in just a few years time, reflecting a major change in the industry.}
\label{fig:timeseries}
\end{figure}

\subsection*{Dependent Variables: Quantifying Innovation}

Quantifying innovation in creative industries often depends on extracting features from unstructured information including music \cite{askin2017makes,wang2019gender}, funding pitches \cite{horvat2018role}, scientific proposals \cite{boudreau2016looking}, video games \cite{de2015game}, and political speeches \cite{barron2018individuals}. These previous works typically compare consistent mathematical descriptions of new products to those that have come before. Outputs that deviate significantly in some way are described as novel. There is a growing body of theoretical and empirical evidence that innovation occurs when content creators combine familiar elements in new ways, which in turn creates possibilities for new combinations \cite{tria2014dynamics,iacopini2018network}. This conceptualization of the ``adjacent possible'' is based on a description of products in terms of comparable constituent elements - in our case the mechanisms of games \cite{kauffman2000investigations}. The literature on ``recombinant innovation'' also frames innovation in terms of new combinations of ingredients \cite{zhang2019recombinant}.

An alternative perspective is to consider the inputs that a product takes. In the case of academic research, the novelty of works cited in a scientific paper can predict breakthrough success \cite{uzzi2013atypical}. Patents also cite prior work - and it is known that patents citing more diverse prior work are generally assigned to more exotic combinations of categories~\cite{aharonson2016mapping,broekel2019using}. We follow the former approach, considering a consistent set of features that describe our database of board games. To quantify innovative outcomes, we will compare a focal board game's combination of mechanisms with those of previously published games. As innovation is a multi-faceted phenomenon, we present several measures to describe the newness of a board game. We compare games in our data with games that were published in previous years using their mechanism vectors. When we are interested in the influence or significance of a game, we compare it with games published in subsequent years. We share a plot of the distributions of all three innovation measures in the Appendix (see Table \ref{fig:dv_dists}).

\subsubsection*{Distinctiveness}
The \textit{distinctiveness} of a game measures the average distance of the game's vector to all games published in the previous 2 years (we replicate our findings with 1 and 5-year windows). We apply the Hamming distance, which simply counts the entries in which the two binary vectors differ. In the case of game mechanisms, the distance adds the number of mechanisms present in game A but not game B to the number of mechanisms present in B but not A. Expressed in mathematical notation, the distinctiveness $D$ of game $g_{i}$ published in year $y$ is defined as:

\begin{gather}
 D(g_{i}) = \frac{\sum_{g \in G_{y-1,y-2}} Hamming(g_{i},g)}{\lvert\{g \in G_{y-1,y-2}\}\rvert},   
\end{gather}

\noindent where $G_{y-1,y-2}$ refers to the set of games published in the previous two years.

\subsubsection*{Novelty}
While distinctiveness measures how atypical a game is, our measure of a game's \textit{novelty} quantifies the extent to which it combines mechanisms in a way completely different from previous games. We calculate the minimum distance of a game to all games published in the previous 2 (resp. 1, 5) years. A game with a novelty score of $k$ has a vector of mechanisms that differs by at least $k$ mechanism from any game the previous window. Though we may mark as novel games that recreate the same combination of a mechanisms as a significantly older game, the revival of such dormant discoveries is an important part of innovation \cite{ferreira2020quantifying}. Expressed in mathematical notation, the novelty $N$ of game $g_{i}$ published in year $y$ is defined as:

\begin{gather}
N(g_{i}) = \min_{g \in G_{y-1,y-2}} Hamming(g_{i},g).
\end{gather}

To simplify our analysis, we map this count variable to a binary variable $N^{0,1}(g_{i})$ taking the value 1 if $N(g_{i})$ is greater than 0, otherwise 0. We do this because we are primarily interested in whether a game implements a new mechanism list or not. The extent of the novelty is to some degree captured by the distinctiveness measure. In any case, our results remain qualitatively unchanged if we model novelty as a count rather than binary outcome.

\subsubsection*{Resonance}
Novel combinations of mechanisms may not be interesting. In the words of Loguidice and Barton, writing about video games, ``if a game does something first, [it doesn't] make it more influential than the later games.'' \cite{loguidice2012vintage} A profound innovation, on the other hand, would inspire future designers to imitate a game. To that end we adapt a measure from Barron et al.'s study of influential speechmaking during the French Revolution \cite{barron2018individuals}. The authors argue that while a speech that differs significantly in content from prior speeches may be novel, it is only influential or \textit{resonant} if future speeches imitate it. A game may implement a surprising combination of mechanisms, but if this combination turns out to be rather strange than interesting, games in the future will not imitate it. The resonance of a game is the difference between its distinctiveness compared to previous games and its distinctiveness compared to subsequent games. 

\begin{gather}
R(g_{i}) = \overbrace{\frac{\sum_{g \in G_{y-1,y-2}} Hamming(g_{i},g)}{\lvert\{g \in G_{y-1,y-2}\}\rvert}}^\text{Distinctiveness relative to past games}
- \overbrace{\frac{\sum_{g \in G_{y+1,y+2}} Hamming(g_{i},g)}{\lvert\{g \in G_{y+1,y+2}\}\rvert}}^\text{Distinctiveness relative to future games}.
\end{gather}

The first term of this expression is simply the distinctiveness of the game $D(g_{i})$, while the second term is its distinctiveness relative to future, rather than past, games. Because the calculation of a games' resonance requires two (resp. 1, 5) years of data after the game was published, we have a smaller sample of games for which we can calculate resonance.

\subsubsection*{The Landscape of Board Game Types}
Our measures of innovation refer to distances between different kinds of games, defined in terms of their vectors of mechanisms. Using this notion of distance between types of games, we can create a map or landscape of the board game industry. In particular we adapt a network visualization method developed by Aharonson and Schilling to study innovation in technology using patent data \cite{aharonson2016mapping} which they call a technological landscape.

\begin{figure}[!b]
\centering
\includegraphics[width=\textwidth]{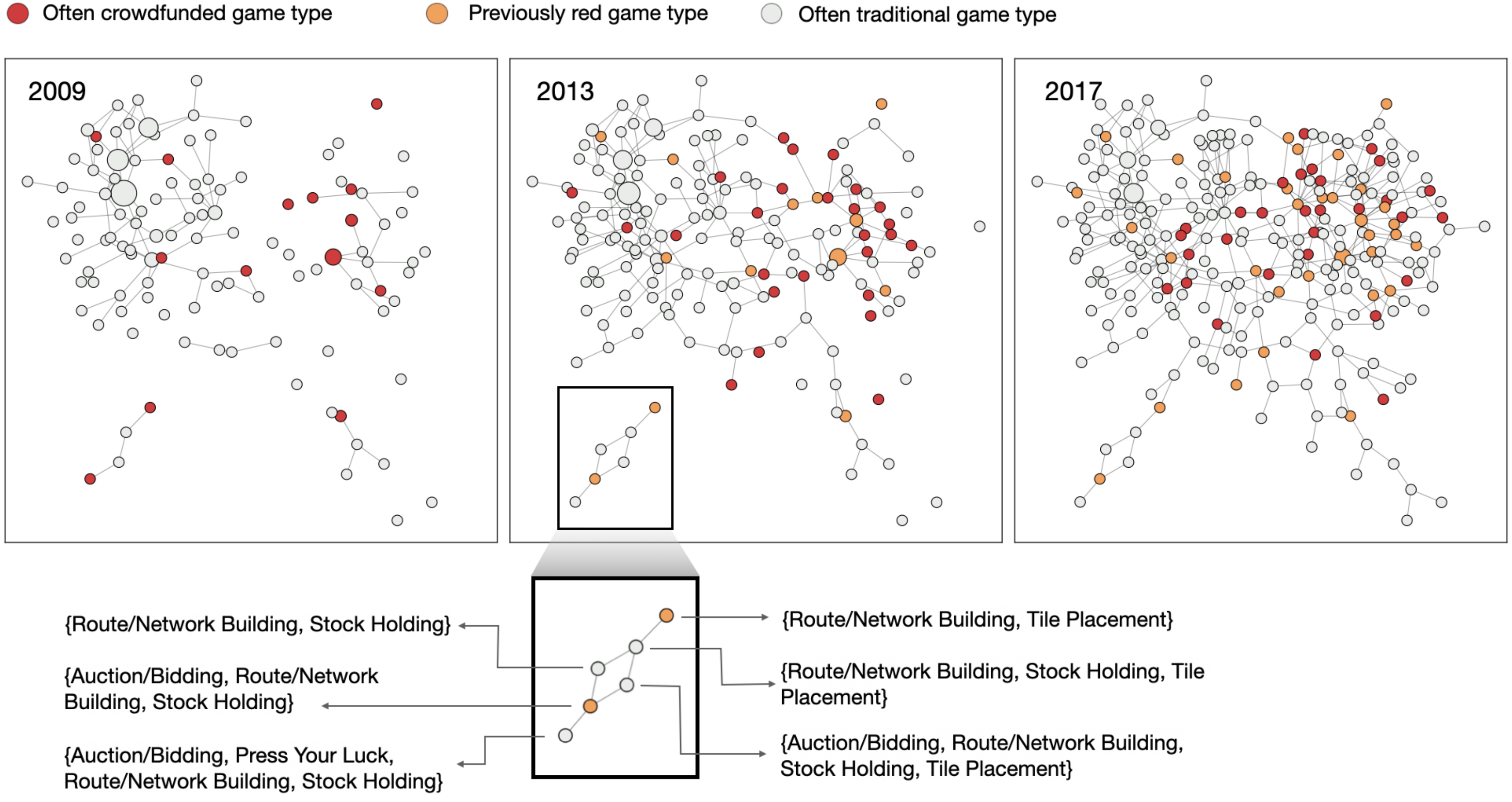}
\captionsetup{width=\linewidth}
\caption{Three snapshots of the networked landscape of board game types. Nodes represent board game types, characterized by their mechanisms. Two nodes are connected if their mechanisms differ in only a single dimension. Node size reflects the number of such games published to dates. Nodes are red if they are often crowdfunded up to that year, grey otherwise. Red nodes turn orange as they are more often implemented by traditionally published games.}
\label{fig:landscape}
\end{figure}

In this network visualization, nodes are unique game vectors that have been realized by a published game. Two nodes are connected by an edge if the Hamming distance of the two vectors is equal to one. Our measures of innovation can be interpreted using the network: a game type is distinct if its vector of mechanism's node is far from the center of the network, away from the other nodes. A game type is novel if it introduces a new node to the network - and moreso if that node is disconnected from the other nodes. Finally, a game type is resonant if subsequently published games are close to it in the network. In this framework, games that implement a new type and are then frequently imitated are the major innovation success stories.

We plot the evolution of the overall board game industry in this network landscape of types in Figure \ref{fig:landscape}. To draw the network we use a physics-inspired force layout algorithm \cite{kamada1989algorithm}. The algorithm simulates the network as a physical system: nodes repel one another as though they were charged particles, while edges act as springs pulling connected nodes together . We run this layout algorithm for the data at the end of 2017, fixing the nodes in previous years to their position. To improve visibility we also filter for game vectors implemented by more than five games in our entire database, and do not plot nodes disconnected from the main component of the network in 2017. Within each snapshot, nodes are larger if there are more games with that vector published by that time. We highlight those nodes for which crowdfunding is particularly common in red. Nodes are colored orange if they were crowdfunding hotspots in previous snapshots, but now tend to be implemented by traditionally funded games.

\begin{figure}[!h]
\centering
\includegraphics[width=\textwidth]{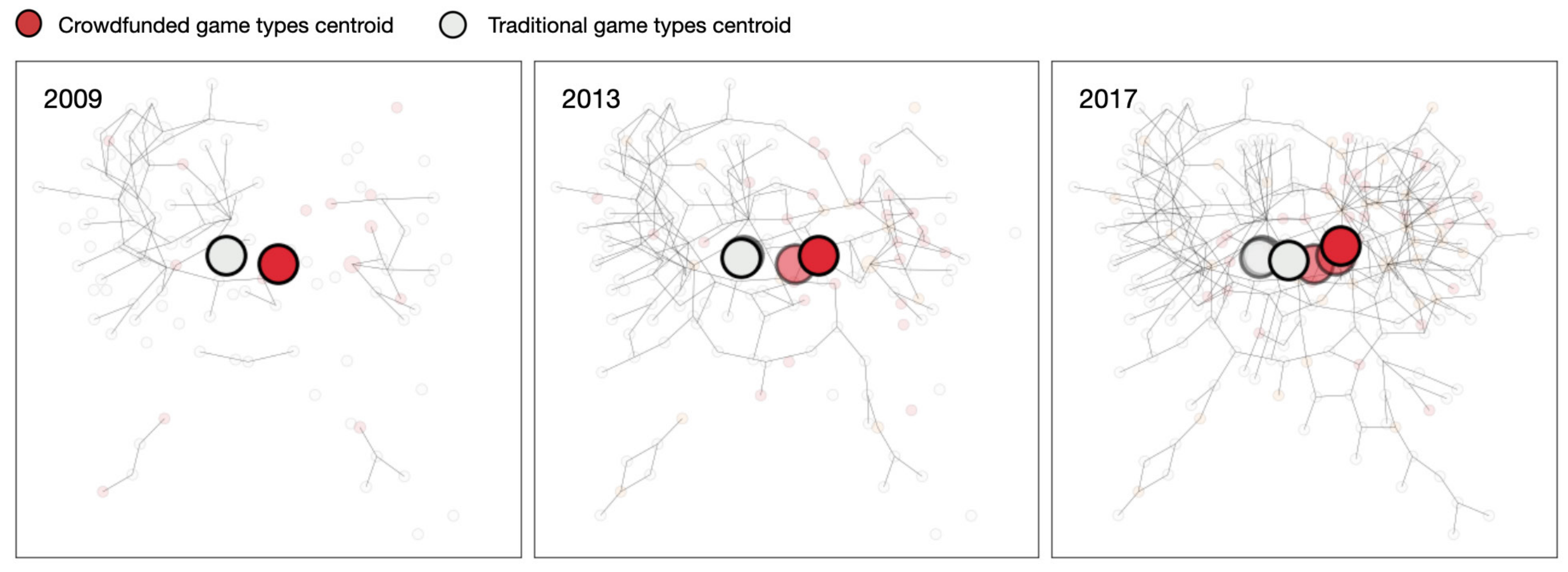}
\captionsetup{width=\linewidth}
\caption{The three board game type landscapes, with the centroids of traditionally published and crowdfunded game types highlighted. Previous period centroids are plotted with transparency to better visual the trend. The crowdfunded game type centroid, in red, moves to the right of the network, and the traditional game type centroid, in grey, follows.}
\label{fig:centroids}
\end{figure}

Qualitatively, we can observe a diversification of game types between 2008 and 2017. Game types that were first implemented by crowdfunded games are often adopted by subsequent traditionally funded games. As whole, we observe the center of gravity of the board game industry shifting dramatically towards territory originally staked out by crowdfunded products. We redraw the three landscape snapshots in Figure~\ref{fig:centroids} comparing the centroid of the game types which tend to be traditionally published with the centroid of the game types tending to be crowdfunded. The traditional centroid seems to follow the crowdfunded one towards the right part of the landscape. In the following section we will provide more rigorous evidence for this visual observation in the form of regression models.

\subsubsection*{Controls}
To exclude potential confounding factors in our regression models estimating the innovation premium of crowdfunded games, we add control terms to our models. First of all, our models also include include year and genre fixed effects to control for overall trends in the game industry and differences between, for example, war games and party games. The genres are: ``strategy'', ``family'', ``wargames'', ``thematic'', ``party'', ``abstract/strategy'', ``childrens'', ``customizable''. We now outline our control variables and discuss how they may confound the relationship between innovation outcomes and crowdfunding.

As discussed in previous sections, crowdfunding enables different kinds of creators to realize their projects. From our descriptive analysis, we know that crowdfunded games are more often made by larger teams of designers and debut designers. Larger teams may, independently of crowdfunding, be more likely to create innovative products because of their greater potential cognitive diversity. Likewise newcomers, regardless of whether they are crowdfunding or not, may bring an outsider perspective. We also expect that newcomers are more likely to be attracted to crowdfunding because of the gatekeeper effect: publishers might prefer to work with proven talent. We therefore control for both team size and whether the design team is making their debut.

There are a number of game attributes that indicate that a game targets a niche market. Niche games might be more innovative only as a function of serving a special market. Niche games might also be more likely crowdfunded, as publishers might stay away from catering to smaller markets. Though we excluded  expansions that do not modify the mechanisms of the original game from our dataset, we have kept non-trivial expansions. Such games may be more likely to be crowdfunded as the designers want to test potential demand for extensions. Expansions may also have limited potential for novelty in general. Similarly, the minimum and maximum number of players a game recommends and its estimated playing time can also highlight niche work: if a designer wants to make a game for dozens of individuals to play together, it may be intrinsically innovative regardless of crowdfunding status. Games targeted at adults may be crowdfunded because traditional publishers hesitate to attach their names to such efforts. \textit{Cards Against Humanity}, one of the most successful crowdfunding projects, not only among board games, is marketed to gamers 18+. We control for these basic board game level features in our models.

Another potential source of bias to our estimates of the relationship between crowdfunded status and innovation outcomes is heterogeneity in complexity between game types. Links between the BGG community and crowdfunded projects suggest that hobby gamers play an important role in the crowdfunding market and may drive the designers to create more complicated games for this group of customers. Traditional publishing houses might be more skeptical of the market for complex games. Game complexity may relate to our innovation measures because complicated games likely implement unfamiliar combinations of mechanisms. We control both for the game's minimum recommended age, as board games designed for younger individuals may be less sophisticated or may emphasize different mechanisms, and BGG's complexity score for each game. BGG users rate the complexity of each game on a 1-5 scale. Inclusion of these controls insures that any relationship we observe between crowdfunding status and innovation outcome is not simply the result of selection into crowdfunding by designers with a preference for complex games.

%Beyond these control terms, we also note two interesting measures of the popularity of games: the number of fans a game has on BGG, and the game's average rating (on a 1 to 10 scale) on BGG. In the following section we will see that these measures indicate that crowdfunded games are significantly more popular on BGG than traditional games. We do not include these terms in our models as the popularity of a game have an unclear relationship with their innovativeness. We prefer to test the influence of crowdfunding using our resonance dependent variable, in part because success on BGG is only one facet of success of a game. 

We report univariate descriptive statistics of all the variables in our dataset in Table \ref{tab:univariate_stats}. Before we introduce the models and our findings we comparing the population of crowdfunded games with traditionally published ones using bivariate statistical tests.

\begin{longtable}{lrrrrrrrr}
\toprule
{} &  Count &     Mean &      Std. &      Min &      25\% &      50\% &      75\% &       Max \\
\midrule
Year Published       &   9170 &  2012.6 &     3.26 &  2006 &  2010 &  2013 &  2015 &   2017 \\
Is Crowdfunded       &   9170 &     0.21 &     0.41 &     0 &     0 &     0 &     0 &      1 \\
Distinctiveness           &   9170 &     5.64 &     1.15 &     3.74 &     4.79 &     5.39 &     6.26 &     18.39 \\
Novelty (Count)           &   9170 &     0.88 &     1 &     0 &     0 &     1 &     1 &     13 \\
Novelty (Binary)     &   9170 &     0.57 &     0.50 &     0 &     0 &     1 &     1 &      1 \\
Resonance         &   7023 &    -0.11 &     0.13 &    -0.52 &    -0.21 &    -0.12 &    -0.02 &      0.45 \\
Is Expansion         &   9170 &     0.12 &     0.33 &     0 &     0 &     0 &     0 &      1 \\
Min. \# of Players          &   9170 &     2.01 &     0.72 &     1 &     2 &     2 &     2 &     10 \\
Max. \# of Players          &   9170 &     4.62 &     2.19 &     0 &     4 &     4 &     6 &     12 \\
Min. Age              &   9170 &     9.67 &     3.98 &     0 &     8 &    10 &    12 &     18 \\
Is Adult/Mature       &   9170 &     0.01 &     0.08 &     0 &     0 &     0 &     0 &      1 \\
Avg. Playing Time (Mins.)     &   9170 &    65 &    73 &     0 &    30 &    45 &    90 &    600 \\
%\# of Users Rated        &   9170 &   681.80 &  2655.93 &    10 &    24 &    72 &   306 &  68885 \\
%\# of Fans                &   9170 &    57.43 &   213.17 &     0 &     3 &    10 &    35 &   5647 \\
%Avg. Rating           &   9170 &     6.78 &     0.88 &     1.43 &     6.23 &     6.81 &     7.38 &      9.94 \\
Avg. Complexity Rating &   9170 &     1.97 &     0.99 &     0.00 &     1.33 &     2.00 &     2.64 &      5.00 \\
Team Size         &   9170 &     1.44 &     0.92 &     1 &     1 &     1 &     2 &     14 \\
Debut               &   9170 &     0.35 &     0.48 &     0 &     0 &     0 &     1 &      1 \\
\bottomrule
\caption{Univariate descriptive statistics of the data, including innovation outcomes, crowdfunding likelihood, game attributes, popularity metrics, perceived complexity.}
\label{tab:univariate_stats}
\end{longtable}

\subsection*{Descriptive findings}
We find several substantive differences between the two group of games, summarized in Table \ref{comparisons}. We report the means of crowdfunded and traditionally published games, and apply a Mann-Whitney U test to measure the statistical significance of their differences. We also report the effect size of a hypothetical classifier distinguishing crowdfunded and traditional games using only the feature in question, measured by the area under the receiver operating characteristic curve (AUC) - in this case equivalent to the common language effect size. 

We find that the teams creating crowdfunded games are slightly larger on average than those making traditional games (3.3 vs 2.9 designers and artists). Larger teams may have greater access to the social capital and networks so essential to crowdfunding \cite{baum2004picking}. While half of published crowdfunded games are made by a debut designer or design team, less than one-third of traditional games are made by newcomers. This suggests that either new entreprenuers are selecting crowdfunding, or are turning to crowdfunding because they cannot tap traditional sources of investment.

Games created via crowdfunding are also different in substantive ways: games for adults (labeled explicitly or having a recommended age of at least 18) are much more likely to be crowdfunded. It may be that some crowdfunded projects have difficulty getting traditional funding for reasons of content. Crowdfunded games are also less likely to be expansions of existing games and have slightly shorter average playing times (roughly 2 minutes, on average).

%Crowdfunded games are also popular: they have nearly twice as many listed ``fans'' on BGG than their traditional counterparts on average. They also have an average ranking .2 points higher (on a 1 to 10 scale). We do not over-interpret this finding to infer that crowdfunded games are on the whole more successful because BGG is a particular community of gaming enthusiasts. While commercial success is likely correlated with these rankings, there are best-selling games that are quite unpopular among the BGG community (such as Monopoly). What we do infer is that the crowdfunded games seem to have robust social networks of hobbyists. 

Turning to our measures of innovation, crowdfunded games have significantly higher distinctiveness, novelty, and resonance than traditional games. However, given the significant potential for confounding factors (as we have seen that crowdfunded games differ in terms of their creators and content), we defer a discussion of the relationship between innovation outcomes and crowdfunding until after we present our models.

\begin{longtable}{lccccccc}

\toprule
             Feature &  Crowdfunded Avg. &  Traditional Avg. &      M-W U &  P Value &  AUC/Effect Size \\
\midrule
           Distinctiveness &              6.05 &              5.53 &  5164212.5 &     <.001 &             0.63 \\
           Novelty (Count) &              1.13 &              0.82 &  5808176.0 &     <.001 &             0.58 \\
     Novelty (Binary) &              0.67 &              0.54 &  5978946.0 &     <.001 &             0.57 \\
         Resonance &             -0.08 &             -0.12 &  4994918.0 &     <.001 &             0.64 \\
         Is Expansion &              0.09 &              0.13 &  6605528.0 &     <.001 &             0.52 \\
 Avg. Complexity Rating &              1.98 &              1.96 &  6765551.5 &     0.05 &             0.51 \\
           %Avg. Rating &              6.94 &              6.74 &  6051399.5 &     <.001 &             0.56 \\
                %\# of Fans &             89.67 &             48.93 &  4047097.5 &     <.001 &             0.71 \\
       Is Adult/Mature &              0.01 &              <.001 &  6885773.0 &     <.001 &             0.50 \\
           Team Size &              1.47 &              1.44 &  6694195.5 &     <.001 &             0.52 \\
               Debut &              0.50 &              0.31 &  5628148.0 &     <.001 &             0.59 \\
\bottomrule
\caption{Comparing the means of key features between crowdfunded and traditionally published games. We apply a Mann-Whitney U test, calculate a p-value and report the effect size via the AUC measure. According to these bivariate tests, crowdfunded games tend to have greater innovation outcomes and bigger fanbases. Their teams are slightly larger and they are more likely to be debut games.}
\label{comparisons}
\end{longtable}

\section*{Models and Results}
We fit three regression models on our dataset, one for each innovation outcome dependent variable. In each model the key independent variable is a binary variable: whether or not the game was crowdfunded. We add controls terms as outlined in the previous section, and fixed effects for both year and genre. Two of our dependent variables are continuous: distinctiveness and resonance. For these variables we fit ordinary least-squares (OLS) regression models. The dependent variable novelty is a binary outcome - either a game implements a never before seen combination of mechanisms or it doesn't. We model this outcome using a logistic regression. 

\linespread{1}
\begin{table}[!htbp] \centering 
\begin{tabular}{@{\extracolsep{5pt}}lccc} 
\\\toprule
 & \multicolumn{3}{c}{\textit{Dependent variable:}} \\ 
\cline{2-4} 
\\[-1.8ex] & Distinctiveness & Novelty & Resonance \\ 
\\[-1.8ex] & \textit{OLS} & \textit{Logistic} & \textit{OLS} \\ 
%\\[-1.8ex] & (1) & (2) & (3)\\ 
 \\
 Is Crowdfunded & 0.235$^{***}$ & 0.412$^{***}$ & 0.014$^{***}$ \\ 
  & (0.032) & (0.061) & (0.004) \\ 
  & & & \\ 
 Team Size & 0.023 & $-$0.025 & 0.003 \\ 
  & (0.012) & (0.025) & (0.001) \\ 
  & & & \\ 
 Debut & 0.049$^{*}$ & 0.362$^{***}$ & 0.003 \\ 
  & (0.024) & (0.048) & (0.003) \\ 
  & & & \\ 
 Avg. Complexity Rating & 0.162$^{***}$ & 0.164$^{***}$ & 0.005$^{*}$ \\ 
  & (0.014) & (0.027) & (0.002) \\ 
  & & & \\ 
 Playing Time (Log Mins) & 0.164$^{***}$ & 0.320$^{***}$ & $-$0.007$^{*}$ \\ 
  & (0.022) & (0.049) & (0.003) \\ 
  & & & \\ 
 Min. \# of Players & $-$0.045$^{*}$ & $-$0.003 & $-$0.009$^{***}$ \\ 
  & (0.018) & (0.034) & (0.002) \\ 
  & & & \\ 
 Max. \# of Players & 0.052$^{***}$ & 0.032$^{**}$ & 0.003$^{***}$ \\ 
  & (0.006) & (0.012) & (0.001) \\ 
  & & & \\ 
 Min. Age & 0.023$^{***}$ & 0.029$^{***}$ & 0.002$^{***}$ \\ 
  & (0.003) & (0.006) & (0.0004) \\ 
  & & & \\ 
 Is Expansion & 0.632$^{***}$ & 0.150$^{*}$ & 0.001 \\ 
  & (0.039) & (0.073) & (0.005) \\ 
  & & & \\ 
 Is Adult/Mature & 0.105 & $-$0.212 & $-$0.014 \\ 
  & (0.178) & (0.289) & (0.016) \\ 
  & & & \\ 
 Constant & 4.071$^{***}$ & $-$1.188$^{***}$ & $-$0.049$^{***}$ \\ 
  & (0.090) & (0.192) & (0.009) \\ 
  & & & \\
\hline \\[-1.8ex] 
Year FE & Yes & Yes & Yes \\ 
Genre FE & Yes & Yes & Yes \\ 
R-Squared & 0.19 & 0.05 & 0.31 \\ 
Observations & 9,170 & 9,170 & 7,023 \\ 
\hline 
 \\[-1.5ex] 
  & \multicolumn{3}{r}{$^{*}$p$<$0.05; $^{**}$p$<$0.01; $^{***}$p$<$0.001} \\ 
\end{tabular} 
  \caption{Model results predicting 2-year windowed measures of innovation. We report McFadden's pseudo-R-squared for the logistic regression model, and robust standard errors for all models.} 
  \label{tab:results} 
\end{table} 
\linespread{1.25}

Our results replicate when modeling novelty as a count variable, measuring the minimum distance of the game's vector to all previously published games, with a Poisson regression (see Table \ref{tab:count_novelty} in the Appendix). As mentioned, we report windowed measures of innovation, looking back (or in the case of resonance, forward) two years. Our results replicate when considering one and five year windows, see Table \ref{tab:1_5_year_robust} in the Appendix. We report robust standard errors, which are marginally higher than the uncorrected errors \cite{angrist2008mostly}. We emphasize that for a range of alternative specifications, filtering or windowing choices, and alternative standard error calculations our models and data tell the same story: controlling for a rich variety of possible confounders, crowdfunded games have significantly higher innovation outcomes.

\begin{figure}[!ht]
\centering
\includegraphics[width=\textwidth]{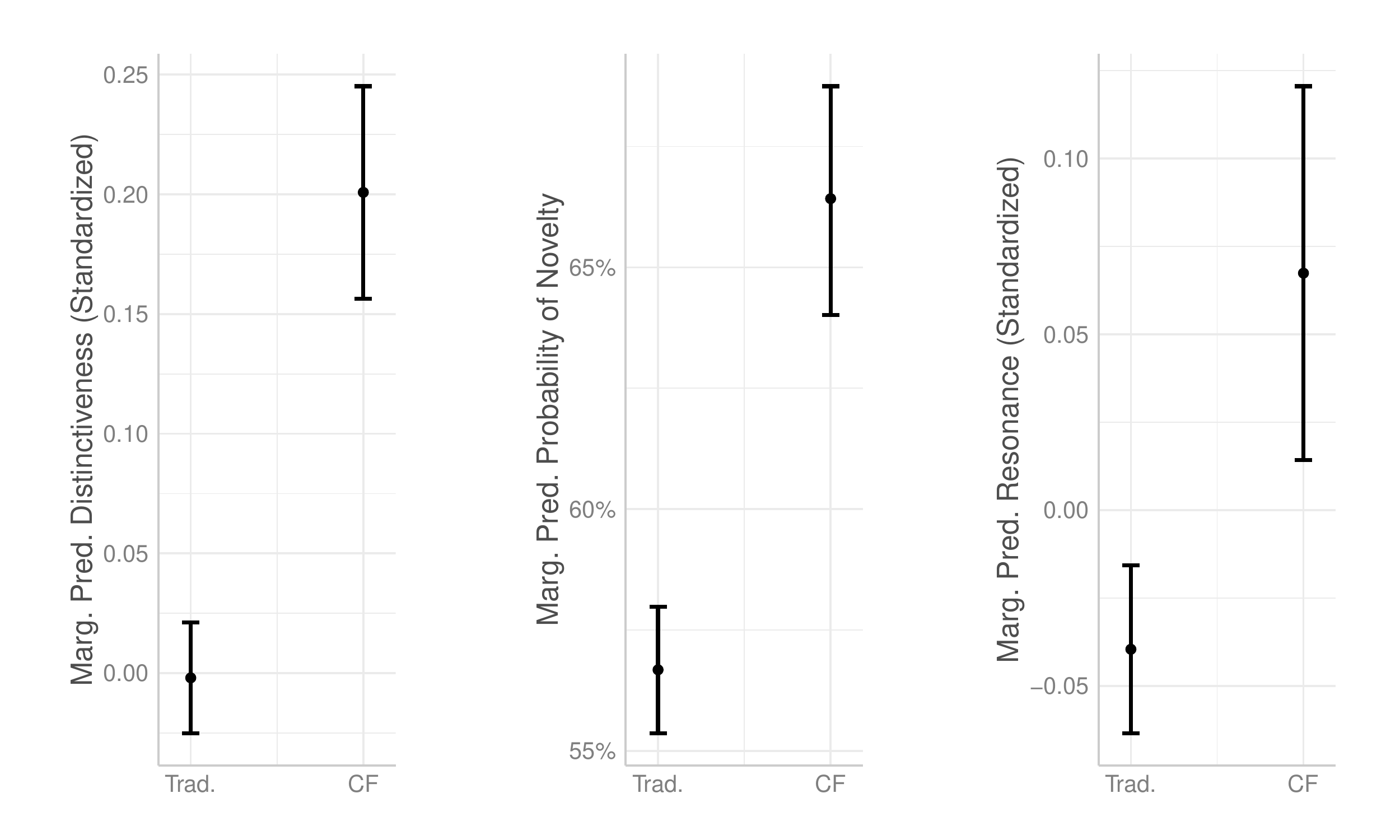}
\caption{Model predicted average innovation outcomes for crowdfunded and traditionally published games. We standardize distinctiveness and resonance to facilitate interpretation. Marginal estimates are derived from the primary models in the text. Bars represent 95\% confidence intervals.}
\label{fig:marg}
\end{figure}

We report the results in Table~\ref{tab:results}. In all three models crowdfunding has a positive and statistically significant relationship with the innovation outcome in question. For instance, a crowdfunded game has 50\% greater odds to implement a novel combination of mechanisms than a similar traditionally published game (exp(.412) $\approx$ 1.5). Crowdfunded games are about .2 standard deviations more distinct (.235/1.15, where 1.15 is the standard deviation of distinctiveness), and about a tenth of a standard deviation more resonant (.014/.14) than traditionally published games. We visually compare the model estimated marginal means for innovation outcomes in Figure \ref{fig:marg}. The estimated effects are calculated using the R package \textit{ggeffects} \cite{ludecke2018ggeffects}. The third model predicts that traditional games will have negative resonance: they have more in common with games created in the past than in the future.

Primary findings aside, we also find relationships of interest between various controls and the dependent variables. Larger teams do not make more innovative games. Debut designers or design teams make more distinct and novel games, though these games are not resonant. This suggests that creating a lasting or influential innovation may require at least some insider experience or legitimization \cite{cattani2008structure}. Previous research suggests that teams that combine insiders and outsiders tend to do well \cite{juhasz2020brokering,vedres2020open}.

Complex games are more innovative, though only slightly more resonant. Longer games tend to be more distinct and novel but actually significantly less resonant. Together these findings highlight a potential tradeoff: a long, complex game may be different from what came before, but overdoing it can make it unlikely that you will be imitated. Matt Leacock, designer of the hit game \textit{Pandemic}, emphasized the importance of making games, quoting Einstein, ``as simple as possible, but no simpler'' \cite{leacockVideo}. The expansion dummy variable also has a different relationship with distinctiveness and novelty than it does with resonance. Expansion games are distinct and novel but not more resonant, suggesting that the industry prefers the original to the imitation. 

\section*{Discussion}
In this paper we have tested a fundamental question about crowdfunding as an institution: does it facilitate the production of innovative products? Despite significant theoretical support in the affirmative, empirical evidence has been lacking - perhaps because it is difficult to source data in which similar crowdfunded and traditionally financed products can be compared. Using a large database of board games sourced from an online community of gaming enthusiasts, we filled this gap and found that crowdfunded games were indeed more innovative in several ways.

To quantify the innovativeness of individual game types determined by their mechanisms, we embedded them in a landscape determined by how they combine fundamental gameplay mechanisms. Different facets of innovation can be quantified in this space using different measures of distance. We said a game is distinct if it has a relatively average distance to previously made games. A game is a novelty if it implements a new combination of mechanisms compared to what came before. On both counts, crowdfunded games scored higher than their traditional counterparts. Lastly, considering subsequent rather than previously published games, we defined a game's resonance as the difference between its distinctiveness in the past versus the future.

In this sense we went beyond our original question and found a surprising result: that crowdfunding not only supports the creation of different products, but that these products predict where the industry is headed. These findings support the notion that crowdfunding represents an extensive disruption \cite{hopp2018disruptive} to the board game industry. Several mechanisms may explain this finding. The crowdfunding process may create stronger ideas through extensive end-user input \cite{von2006democratizing}. The open and public nature of crowdfunding may facilitate knowledge spillovers \cite{jones2020knowledge}. Future work is needed to understand the relative importance of these factors.

We can also point to specific interesting counterfactuals that we cannot test with the data at hand but that would illuminate the relationship between crowdfunding and innovation. We do not observe failed projects on either the crowdfunded or traditionally published sides, and cannot observe differences in preferences between crowd and expert funders, which are thought to be significant \cite{song2019mining}. Randomized control trials, though expensive, and natural experiments, though rare, would present a significant complement to our observational study.

Even with our results, we caution that crowdfunding is not a silver bullet for the problem of sluggish innovation in the advanced economies. Indeed, despite much fanfare, crowdfunding still makes up a small share of investment in the economy. Perhaps most important is the question of scale: while some coordination is required to create a new board game, board games have relatively low capital requirements and can be made by small teams or individuals. Can crowdfunding effectively contribute to more organizationally and capital-intensive innovation? 

Perhaps it is better to ask which sectors are becoming more amenable to crowdfunding. Trends in digitalization and nascent fields like additive manufacturing are decreasing the cost of communication, coordination, and prototyping. We argue that these forces are expanding the industries and technological areas to which our findings apply.  Crowdfunding likely has significant potential to support innovation in non-traditional settings such as DIY labs \cite{galvin2020leveraging}, makerspaces \cite{halbinger2018role,davies2017hackerspaces}, and open source software \cite{overney2020not}. It may also smooth out some of the strong geographical economic forces that concentrate investment in major hubs \cite{breznitz2020crowdfunding}. Future work to measure and understand the innovation impact of crowdfunding should focus on this expanding frontier of places, sectors, and people where it can make a difference.

\vspace{-.4cm}
\section*{Acknowledgements}
\vspace{-.4cm}
We thank S\'andor Juh\'asz, Hannes Fernstr\"om, Tamer Khraisha, Gerg\H{o} T\'oth, David Deritei, Mil\'an Janosov, Anna May, Theresa Gessler, the participants of seminars at the Chair of Computational Social Sciences and Humanities at RWTH Aachen University, and the Institute for Analytical Sociology at Link\"oping University for helpful comments and suggestions. We are grateful to Zs\'ofia Cz\'em\'an for assistance with our figures.

\vspace{-.4cm}
\section*{Conflict of interest}
\vspace{-.4cm}
The authors declare that they have no conflict of interest.

\bibliographystyle{spmpsci}       % APS-like style for physics
\bibliography{bibliography.bib}   % name your BibTeX data base

\clearpage

\section*{Appendix}

In the following table (Table \ref{Appendix:mechanism_table}), we list the game mechanism typology in our dataset, sourced from BoardGameGeek's website in 2017. We also display the platform's list of popular games implementing each particular mechanism. 
\begin{longtable}{| p{.2\textwidth} | p{.6\textwidth} |} 
%\resizebox{\textwidth}{!}{%
\toprule
 \textbf{Mechanism}  & \textbf{Examples}                                                                    \\ \midrule
\multicolumn{1}{|l|}{Acting}                        & \multicolumn{1}{l|}{Times Up!}                                                                 \\ \midrule
\multicolumn{1}{|l|}{Action/Movement Programming}   & \multicolumn{1}{l|}{Shogun, Robo   Rally}                                                      \\ \midrule
\multicolumn{1}{|l|}{Action Point Allowance System} & \multicolumn{1}{l|}{Tikal}                                                                     \\ \midrule
\multicolumn{1}{|l|}{Area control/Area influence}   & \multicolumn{1}{l|}{Eclipse}                                                                   \\ \midrule
\multicolumn{1}{|l|}{Area Enclosure}                & \multicolumn{1}{l|}{Go, Boxes}                                                                 \\ \midrule
\multicolumn{1}{|l|}{Area Movement}                 & \multicolumn{1}{l|}{El Grande,   Dead of Winter, War of the Ring}                              \\ \midrule
\multicolumn{1}{|l|}{Area-Impulse}                & \multicolumn{1}{l|}{Storm over   Arnhem, Thunder at Cassino, Turning Point: Stalingrad} \\ \midrule
\multicolumn{1}{|l|}{Auction/Bidding}               & \multicolumn{1}{l|}{Modern Art,   Ra}                                                          \\ \midrule
\multicolumn{1}{|l|}{Betting/Wagering}              & \multicolumn{1}{l|}{Tichu}                                                                     \\ \midrule
\multicolumn{1}{|l|}{Campaign/Battle Card Driven} & \multicolumn{1}{l|}{We the People, Hannibal: Rome vs Carthage, Twilight Struggle}                         \\ \midrule
\multicolumn{1}{|l|}{Card Drafting}                 & \multicolumn{1}{l|}{Through the Ages, Mage Knight}                                           \\ \midrule
\multicolumn{1}{|l|}{Chit-Pull System}              & \multicolumn{1}{l|}{Zulus on the Ramparts, A Victory Lost}                                   \\ \midrule
\multicolumn{1}{|l|}{Co-operative Play}             & \multicolumn{1}{l|}{Pandemic,   Escape, Battlestar Galactica, Castle Panic}                    \\ \midrule
\multicolumn{1}{|l|}{Commodity Speculation}         & \multicolumn{1}{l|}{Acquire,   Kanban, Merchant of Venus}                                      \\ \midrule
\multicolumn{1}{|l|}{Crayon Rail System}            & \multicolumn{1}{l|}{Empire  Builder, Eurorails}                                               \\ \midrule
\multicolumn{1}{|l|}{Deck/Pool Building}            & \multicolumn{1}{l|}{Dominion,   Star Realms, Marvel Legendary}                                 \\ \midrule
\multicolumn{1}{|l|}{Dice Rolling}                  & \multicolumn{1}{l|}{Roll for the Galaxy, Mice and Mystics, Yahtzee}                          \\ \midrule
\multicolumn{1}{|l|}{Grid movement}                 & \multicolumn{1}{l|}{Space Hulk, Forbidden Desert}                                            \\ \midrule
\multicolumn{1}{|l|}{Hand Management}               & \multicolumn{1}{l|}{Android: Netrunner, Through the Ages}                                    \\ \midrule
\multicolumn{1}{|l|}{Hex-and-Counter}               & \multicolumn{1}{l|}{Twilight Imperium, Advanced Squad Leader,}                               \\ \midrule
\multicolumn{1}{|l|}{Line Drawing}                  & \multicolumn{1}{l|}{Telestrations, Pictionary, Cranium, Sprouts}                             \\ \midrule
\multicolumn{1}{|l|}{Memory}                        & \multicolumn{1}{l|}{Codenames, Hanabi, Coup, Sleuth, Clue}                                   \\ \midrule
\multicolumn{1}{|l|}{Modular Board}                 & \multicolumn{1}{l|}{Settlers of Catan, Mansions of Madness, Blue Moon City} \\ \midrule
\multicolumn{1}{|l|}{Paper-and-Pencil}              & \multicolumn{1}{l|}{Eat Poop You Cat, Scattergories}                                         \\ \midrule
\multicolumn{1}{|l|}{Partnerships}                  & \multicolumn{1}{l|}{The Resistance, Dune, Tragedy Looper, Ultimate Werewolf}                 \\ \midrule
\multicolumn{1}{|l|}{Pattern Building}              & \multicolumn{1}{l|}{Castles of Mad King Ludwig}                                              \\ \midrule
\multicolumn{1}{|l|}{Pattern Recognition}           & \multicolumn{1}{l|}{Quirkle,   Ingenious, Ubongo, Jungle Speed, Cathedral, Pentagon}           \\ \midrule
\multicolumn{1}{|l|}{Pick-up and Deliver}           & \multicolumn{1}{l|}{Merchants and Marauders, Indonesia, Genoa}                               \\ \midrule
\multicolumn{1}{|l|}{Player elimination}            & \multicolumn{1}{l|}{Magic: The Gathering, King of Tokyo, Diplomacy}                          \\ \midrule
\multicolumn{1}{|l|}{Point to Point Movement}       & \multicolumn{1}{l|}{Arkham Horror, Tales of the Arabian Nights}                              \\ \midrule
\multicolumn{1}{|l|}{Press your Luck}               & \multicolumn{1}{l|}{Goa, Ra, Incan Gold}                                                     \\ \midrule
\multicolumn{1}{|l|}{Rock-Paper-Scissors}           & \multicolumn{1}{l|}{Sid Meier's Civilization, Yomi, Dungeon Quest}                           \\ \midrule
\multicolumn{1}{|l|}{Role Playing}                  & \multicolumn{1}{l|}{Chaos in the old world, Descent: Journeys in the Dark}                   \\ \midrule
\multicolumn{1}{|l|}{Roll/Spin and Move}            & \multicolumn{1}{l|}{Monopoly, Marrakech, Formula Dè}                                         \\ \midrule
\multicolumn{1}{|l|}{Route/Network building}        & \multicolumn{1}{l|}{Power Grid, Railways of the World, Food Chain Magnate}                   \\ \midrule
\multicolumn{1}{|l|}{Secret Unit Deployment}        & \multicolumn{1}{l|}{Letters from Whitechapel, Fury of Dracula}                               \\ \midrule
\multicolumn{1}{|l|}{Set Collection}                & \multicolumn{1}{l|}{Gin Rummy, Lords of Waterdeep}                                           \\ \midrule
\multicolumn{1}{|l|}{Simulation}                  & \multicolumn{1}{l|}{Castles of Burgundy, The Voyages of Marco Polo, Flash Point: Fire Rescue} \\ \midrule
\multicolumn{1}{|l|}{Simultaneous Action Selection} & \multicolumn{1}{l|}{7 Wonders}                                                                 \\ \midrule
\multicolumn{1}{|l|}{Singing}                       & \multicolumn{1}{l|}{Cranium}                                                                   \\ \midrule
\multicolumn{1}{|l|}{Stock Holding}                 & \multicolumn{1}{l|}{Mombasa, Rapa Nui, Tesla vs Edison}                                      \\ \midrule
\multicolumn{1}{|l|}{Storytelling}                  & \multicolumn{1}{l|}{Tales of the Arabian Nights, Dixit, Rory's Story Cubes}                  \\ \midrule
\multicolumn{1}{|l|}{Take That}                     & \multicolumn{1}{l|}{Cosmic Encounter, Saboteur, Bang, Munchkin}                              \\ \midrule
\multicolumn{1}{|l|}{Tile Placement}                & \multicolumn{1}{l|}{Carcassonne}                                                               \\ \midrule
\multicolumn{1}{|l|}{Time Track}                    & \multicolumn{1}{l|}{Horus Heresy}                                                              \\ \midrule
\multicolumn{1}{|l|}{Trading}                       & \multicolumn{1}{l|}{Tikal,   Antiquity, Fief, Bohnanza}                                        \\ \midrule
\multicolumn{1}{|l|}{Trick-taking}                  & \multicolumn{1}{l|}{Hearts,  Haggis, Rook}                                                    \\ \midrule
\multicolumn{1}{|l|}{Variable Phase Order}          & \multicolumn{1}{l|}{Puerto Rico,   Race for the Galaxy, Citadels, Myth}                        \\ \midrule
\multicolumn{1}{|l|}{Variable Player Powers}        & \multicolumn{1}{l|}{Cosmic Encounter, Dominant Species, Sentinel of the Multiverse}          \\ \midrule
\multicolumn{1}{|l|}{Voting}                        & \multicolumn{1}{l|}{Werewolf, Battlestar Galactica, Patchistory}                             \\ \midrule
\multicolumn{1}{|l|}{Worker Placement}              & 
\multicolumn{1}{l|} {Agricola, Terra Mystica, Caylus, Caverna, Le Havre, Orleans}                      \\ \bottomrule
\caption{Table of mechanisms in games, with examples. Note we use the mechanism system as defined in 2017. Source: \url{https://web.archive.org/web/20170722061312/https://boardgamegeek.com/wiki/page/mechanism}}
\label{Appendix:mechanism_table}

\end{longtable}

\begin{figure}[!h]
    \centering
    \includegraphics[width=\textwidth]{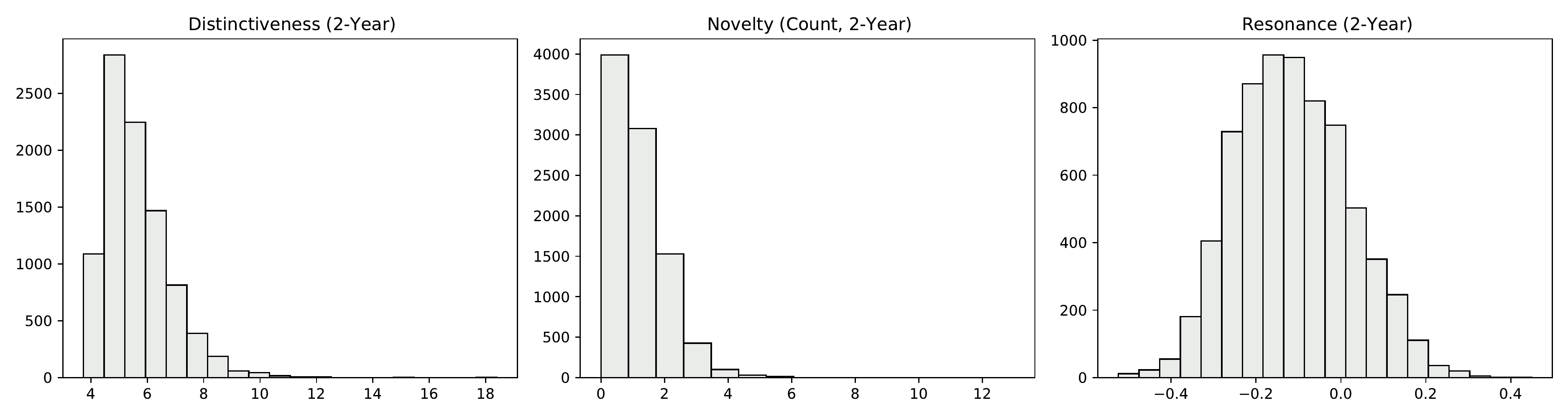}
    \caption{Distributions of dependent variables.}
    \label{fig:dv_dists}
\end{figure}
\newpage

\section*{Robustness Tests}

\begin{table}[!htbp] \centering 
\begin{tabular}{@{\extracolsep{5pt}}lcccccc} 
\\[-1.8ex]\hline 
\hline \\[-1.8ex] 
 & \multicolumn{6}{c}{\textit{Dependent variable:}} \\ 
\cline{2-7} 
\\[-1.8ex] & 1Y-Distinct. & 1Y-Nov. & 1Y-Res. & 5Y-Distinct. & 5Y-Nov. & 5Y-Res. \\ 
\\[-1.8ex] & \textit{OLS} & \textit{Logistic} & \textit{OLS} & \textit{OLS} & \textit{Logistic} & \textit{OLS} \\ 
 \\[-1.8ex] 
 Is Crowdfunded & 0.237$^{***}$ & 0.445$^{***}$ & 0.010$^{**}$ & 0.242$^{***}$ & 0.365$^{***}$ & 0.055$^{***}$ \\ 
  & (0.032) & (0.064) & (0.003) & (0.033) & (0.059) & (0.012) \\ 
  & & & & & & \\ 
 Team Size & 0.022 & $-$0.046 & $-$0.0001 & 0.027$^{*}$ & $-$0.009 & 0.004 \\ 
  & (0.012) & (0.025) & (0.001) & (0.012) & (0.024) & (0.003) \\ 
  & & & & & & \\ 
 Debut & 0.049$^{*}$ & 0.289$^{***}$ & 0.002 & 0.050$^{*}$ & 0.365$^{***}$ & 0.002 \\ 
  & (0.024) & (0.050) & (0.002) & (0.024) & (0.047) & (0.005) \\ 
  & & & & & & \\ 
 Avg. Complexity Rating & 0.163$^{***}$ & 0.172$^{***}$ & 0.005$^{**}$ & 0.162$^{***}$ & 0.167$^{***}$ & $-$0.001 \\ 
  & (0.014) & (0.028) & (0.002) & (0.014) & (0.027) & (0.003) \\ 
  & & & & & & \\ 
 Playing Time (log mins.) & 0.166$^{***}$ & 0.320$^{***}$ & $-$0.006$^{*}$ & 0.162$^{***}$ & 0.293$^{***}$ & $-$0.002 \\ 
  & (0.022) & (0.048) & (0.003) & (0.022) & (0.050) & (0.005) \\ 
  & & & & & & \\ 
 Min. \# of Players & $-$0.044$^{*}$ & 0.053 & $-$0.005$^{**}$ & $-$0.049$^{**}$ & $-$0.029 & $-$0.014$^{***}$ \\ 
  & (0.018) & (0.036) & (0.002) & (0.019) & (0.034) & (0.003) \\ 
  & & & & & & \\ 
 Max. \# of Players & 0.052$^{***}$ & 0.030$^{*}$ & 0.003$^{**}$ & 0.052$^{***}$ & 0.040$^{***}$ & 0.002$^{*}$ \\ 
  & (0.006) & (0.012) & (0.001) & (0.006) & (0.012) & (0.001) \\ 
  & & & & & & \\ 
 Min. Age & 0.023$^{***}$ & 0.025$^{***}$ & 0.001$^{***}$ & 0.024$^{***}$ & 0.038$^{***}$ & 0.003$^{***}$ \\ 
  & (0.003) & (0.006) & (0.0003) & (0.003) & (0.006) & (0.001) \\ 
  & & & & & & \\ 
 Is Expansion & 0.616$^{***}$ & 0.119 & $-$0.005 & 0.654$^{***}$ & 0.278$^{***}$ & 0.010 \\ 
  & (0.039) & (0.075) & (0.005) & (0.039) & (0.072) & (0.008) \\ 
  & & & & & & \\ 
 Is Adult/Mature & 0.103 & $-$0.379 & $-$0.021 & 0.115 & 0.080 & 0.057 \\ 
  & (0.177) & (0.295) & (0.014) & (0.180) & (0.288) & (0.041) \\ 
  & & & & & & \\ 
 Constant & 4.083$^{***}$ & $-$0.980$^{***}$ & $-$0.020$^{*}$ & 4.047$^{***}$ & $-$1.669$^{***}$ & $-$0.183$^{***}$ \\ 
  & (0.090) & (0.196) & (0.008) & (0.091) & (0.194) & (0.014) \\ 
  & & & & & & \\ 
\hline \\[-1.8ex] 
Year FE & Yes & Yes & Yes & Yes & Yes & Yes \\ 
Genre FE & Yes & Yes & Yes & Yes & Yes & Yes \\ 
R-squared & 0.19 & 0.04 & 0.39 & 0.19 & 0.05 & 0.26 \\ 
\hline 
\hline \\[-1.8ex] 
 & \multicolumn{6}{r}{$^{*}$p$<$0.05; $^{**}$p$<$0.01; $^{***}$p$<$0.001} \\ 
\end{tabular} 
  \caption{Robustness to alternative year ranges for innovation measures. One and five year winnowed innovation measures: distinctiveness, novelty, and resonance. These models replicate our primary findings with similar effect sizes and statistical significance: that crowdfunded games are more innovative. Our primary models in the text have two-year measures as dependent variables. We report McFadden's pseudo-R-squared for the logistic regression models, and robust standard errors for all models.} 
  \label{tab:1_5_year_robust} 
\end{table}

\begin{table}[!htbp] \centering 
\begin{tabular}{@{\extracolsep{5pt}}lccc} 
\\[-1.8ex]\hline 
\hline \\[-1.8ex] 
 & \multicolumn{3}{c}{\textit{Dependent variable:}} \\ 
\cline{2-4} 
\\[-1.8ex] & 1-Year Count Novelty &2-Year Count Novelty& 5-Year Count Novelty \\ 
\\[-1.8ex] & \multicolumn{3}{c}{\textit{Poisson Regressions}}\\ 
\hline \\[-1.8ex] 
 Is Crowdfunded & 0.211$^{***}$ & 0.226$^{***}$ & 0.243$^{***}$ \\ 
  & (0.026) & (0.029) & (0.033) \\ 
  & & & \\ 
 Team Size & $-$0.014 & $-$0.009 & $-$0.0004 \\ 
  & (0.011) & (0.013) & (0.014) \\ 
  & & & \\ 
 Debut & 0.130$^{***}$ & 0.172$^{***}$ & 0.199$^{***}$ \\ 
  & (0.021) & (0.024) & (0.027) \\ 
  & & & \\ 
 Avg. Complexity Rating & 0.098$^{***}$ & 0.102$^{***}$ & 0.117$^{***}$ \\ 
  & (0.013) & (0.015) & (0.017) \\ 
  & & & \\ 
 Playing Time (log mins.) & 0.196$^{***}$ & 0.225$^{***}$ & 0.231$^{***}$ \\ 
  & (0.027) & (0.032) & (0.035) \\ 
  & & & \\ 
 Min. \# of Players & $-$0.010 & $-$0.029 & $-$0.054$^{**}$ \\ 
  & (0.016) & (0.018) & (0.020) \\ 
  & & & \\ 
 Max. \# of Players & 0.021$^{***}$ & 0.026$^{***}$ & 0.034$^{***}$ \\ 
  & (0.006) & (0.006) & (0.007) \\ 
  & & & \\ 
 Min. Age & 0.016$^{***}$ & 0.019$^{***}$ & 0.026$^{***}$ \\ 
  & (0.003) & (0.004) & (0.004) \\ 
  & & & \\ 
 Is Expansion & 0.079$^{*}$ & 0.072 & 0.125$^{**}$ \\ 
  & (0.034) & (0.037) & (0.041) \\ 
  & & & \\ 
 Is Adult/Mature & 0.001 & 0.008 & 0.102 \\ 
  & (0.124) & (0.136) & (0.135) \\ 
  & & & \\ 
 Constant & $-$0.756$^{***}$ & $-$1.027$^{***}$ & $-$1.460$^{***}$ \\ 
  & (0.094) & (0.108) & (0.123) \\ 
  & & & \\ 
\hline \\[-1.8ex] 
Year FE & Yes & Yes & Yes \\ 
Genre FE & Yes & Yes & Yes \\ 
R-squared & 0.03 & 0.04 & 0.04 \\ 
\hline 
\hline \\[-1.8ex] 
 & \multicolumn{3}{r}{$^{*}$p$<$0.05; $^{**}$p$<$0.01; $^{***}$p$<$0.001} \\ 
\end{tabular} 
  \caption{Poisson models predicting count novelty calculated with 1, 2, and 5 year look-back. This alternative formulation of innovation for the dependent variable aligns with the findings we report in the text: crowdfunded games are significantly more likely to implement a novel combination of mechanisms. We report McFadden's pseudo-R-squared and robust standard errors for all three models.} 
  \label{tab:count_novelty} 
\end{table} 

\end{document}